\documentclass[pra,showpacs]{revtex4} \usepackage{latexsym}
\usepackage{amsfonts} \usepackage{amsmath} 

\newcommand{\kt}[1]{\ensuremath{|#1\rangle}}
\newcommand{\br}[1]{\ensuremath {\langle #1|}}
\newcommand{\bk}[2]{\ensuremath {\langle #1|#2 \rangle}}

\newcommand{\wig}{\ensuremath p\,\chi [\ms\,j]}
\newcommand{\wigh}{\ensuremath p\,\lambda [\ms\,j]}
\newcommand{\wigx}{\ensuremath p\,\xi [\ms\,j]}
\newcommand{\vp}{{\bf p}}

\newcommand{\mmp}{\mathsf{p}}

\newcommand{\mk}{\mathsf{k}}
\newcommand{\vP}{{\bf P}}

\newcommand{\ms}{\mathsf{s}}

\newcommand{\slg}{{\rm SL}(2, \mathbb{C})}
\newcommand{\su}{{\rm SU}(2)}
\newcommand{\ptr}{\tilde{\mathcal{P}}^\uparrow_+}
\newcommand{\pt}{\tilde{\mathcal{P}}}
\newcommand{\HS}{{\mathcal{H}}}

\newcommand{\tl}{\tilde{\lambda}}
\newcommand{\la}{\lambda}

\begin{document}

\title{Basis states for relativistic dynamically entangled particles}

\author{N.L.~Harshman}
\affiliation{Department of Computer Science, Audio Technology and Physics\\
4400 Massachusetts Ave., NW \\ American University\\ Washington, DC 20016-8058}

\begin{abstract}

In several recent papers on entanglement in relativistic quantum systems and relativistic Bell's inequalities, relativistic Bell-type two-particle states have been constructed in analogy to non-relativistic states.  These constructions do not have the form suggested by relativistic invariance of the dynamics.  Two relativistic formulations of Bell-type states are shown for massive particles, one using the standard Wigner spin basis and one using the helicity basis.  The construction hinges on the use of Clebsch-Gordan coefficients of the Poincar\'e group to reduce the direct product of two unitary irreducible representations (UIRs) into a direct sum of UIRs.  
\end{abstract}

\pacs{03.67.Mn,03.65.Ud, 11.80.Et}

\maketitle

\section{Introduction}

Relativistic quantum information theory is a field of growing interest (see the review \cite{peres04}). 
Tracing back to Bell's famous re-imagining of the Einstein-Podolosky-Rosen paradox, a standard system of interest is two particles with spins entangled due to their production in the decay or scattering.  Various authors have considered the entanglement of two relativistic particles~\cite{czachor97, alsing02,gingrich02,ahn03,ahnarch03,pachos03,terashima03,bergou03,soo04,kim04}.
In this paper, we present a general scheme for constructing relativistic entangled bases that respects the Poincar\'e invariance of the dynamics that produces the entanglement.

In analogy to the non-relativistic case~\cite{neilson}, the authors of \cite{czachor97,alsing02,ahn03,ahnarch03,terashima03,soo04,kim04} consider a basis for ``EPR states'' or ``Bell-type states'' for two relativistic spin 1/2 particles with the form of one or more of the following states:
\begin{eqnarray}\label{crap}
\kt{\psi_{00}} &=& \frac{1}{\sqrt{2}}\left(\kt{\vp_1, \uparrow}\otimes\kt{\vp_2, \uparrow} + 
\kt{\vp_1, \downarrow}\otimes\kt{\vp_2, \downarrow}\right)\nonumber\\
\kt{\psi_{01}} &=& \frac{1}{\sqrt{2}}\left(\kt{\vp_1, \uparrow}\otimes\kt{\vp_2, \downarrow} + 
\kt{\vp_1, \downarrow}\otimes\kt{\vp_2, \uparrow}\right)\nonumber\\
\kt{\psi_{10}} &=& \frac{1}{\sqrt{2}}\left(\kt{\vp_1, \uparrow}\otimes\kt{\vp_2, \uparrow} - 
\kt{\vp_1, \downarrow}\otimes\kt{\vp_2, \downarrow}\right)\nonumber\\
\kt{\psi_{11}} &=& \frac{1}{\sqrt{2}}\left(\kt{\vp_1, \uparrow}\otimes\kt{\vp_2, \downarrow} - 
\kt{\vp_1, \downarrow}\otimes\kt{\vp_2, \uparrow}\right),
\end{eqnarray}
where the 3-momentum is considered in the center-of-mass (COM) frame $\vp_1=-\vp_2$.
Since in the non-relativistic case, the spin degrees of freedom and momentum degrees of freedom are independent, states like (\ref{crap}) are useful in the analysis of dynamically-entangled states far from the scattering or decay region because the interaction will be diagonal in total intrinsic spin $s$.  Then the first three states form a basis for the $s=1$ triplet subspace in the COM frame and the last one corresponds to the $s=0$ singlet COM subspace.  Even though the states (\ref{crap}) are not eigenstates of the total angular momentum, their transformation properties under rotations are still straightforward (see below).

However, in the relativistic case the separation of the angular momentum of a particle between orbital and spin is frame-dependent and the value of the spin-component depends on the momentum of the particle.  In other words, Poincar\'e covariant dynamics require the dynamics to be diagonal in the total COM energy squared $\ms$ and the total angular momentum $j$~\cite{weinberg}.  Interactions are diagonal in $j$, and so any particular scattering or decay channel will be in a partial wave specified by $j$.  Therefore, the basis vectors of the partial waves are an alternate choice of bases for relativistic entangled states that has a closer connection to the physical dynamics that cause the entanglement.

Several of these authors give meaning to $\kt{\vp_1, \uparrow}$ in (\ref{crap}) as an element of the non-unitary spinor representations of the Poincar\'e group.  Others work with $\kt{\vp_1, \uparrow}$ as an (singular, non-normalizable) element of a unitary irreducible representation (UIR) of the restricted Poincar\'e group $\ptr$.  In this paper, we will work with UIRs of $\ptr$.  Then the process of reducing direct product states like (\ref{crap}) to partial waves is carried out by finding the Clebsch-Gordan coefficients (CGCs) for $\ptr$.  In general, direct products of UIRs are reducible, a common example being the direct product of two UIRs $\HS(j_1)$ and $\HS(j_2)$ of the rotation group being decomposed into a direct sum of UIRs:
\begin{equation}
\HS(j_1)\otimes\HS(j_2) = \bigoplus_{j=|j_1 - j_2|}^{j_1 + j_2} \HS(j)_{j_1 j_2},
\end{equation}
where $\HS(j)_{j_1 j_2}$ is labeled by the invariants of the original UIRs.  The CGCs for the rotation group then show how to expand direct product vectors in terms of direct sum vectors and vice versa.
Although more complicated for the non-compact group $\ptr$, similar decompositions are possible.  The basis vectors for the UIRs in the direct sum spaces, re-expressed using the CGCs, form a natural basis for describing entanglement produced by Poincar\'e invariant dynamics.

In what follows, the necessary details of single particle UIRs will be briefly reviewed.  Then an overview of Clebsch-Gordan coefficients (CGCs) and the reduction problem for $\ptr$ will be given, followed by the explicit form of the CGCs using spin-orbit coupling in the Wigner spin basis and using helicity coupling the helicty basis.  As an example, the last section will explicitly show the two basis vectors corresponding to the center-of-mass frame of a $j=0$ composition, and the twelve center-of-mass frame basis vectors of the four direct sum UIRs with $j=1$.

\section{One-Particle Relativistic States}

The results in this section are well-known and trace back to Wigner's seminal work~\cite{wigner39} on classifying the unitary irreducible representations of the Poincar\'e group.  We review these results briefly to clarify notation, to focus on the definition of the spin-component operator, and to correct mistaken assumptions made in \cite{gingrich02,ahn03,ahnarch03,kim04}.

Here we work exclusively with the proper, orthochronous, quantum-mechanical (i.e., projective) Poincar\'e group $\ptr$ which is isomorphic to the semidirect product $\slg + \mathbb{R}^4$.  We will denote these elements as $(\alpha, a)$, where $\alpha\in{\rm SL}(2, \mathbb{C})$ and $a\in\mathbb{R}^4$.  The group multiplication law is 
\begin{equation}
(\alpha', a')(\alpha, a) = (\alpha'\alpha, a' + \Lambda(\alpha') a),
\end{equation}
where $\Lambda(\alpha)\in{\rm SO}(1,3)$ is a well-known two-to-one homomorphism~\cite{ruhl}.  The subset ${\rm SU}(2)\subset{\rm SL}(2, \mathbb{C})$ corresponds to the rotation subgroup of transformations; $a\in\mathbb{R}^4$ is the translation subgroup.

Representations of $\ptr$ are constructed using the method of induced representations, first carried out by Wigner~\cite{wigner39} and generalized by Mackey~\cite{mackey}.  The technique relies on building the representation for the full group $\ptr$ from representations of a subgroup, for massive representations typically chosen to be $H = \su \times \mathbb{R}^4$ (see \cite{ruhl,schaaf,weinberg} for details).  Here we only consider positive energy representations.

%The $\tilde{\mathcal{P}}^\uparrow_+$ is a 10-parameter Lie group whose generators fulfill the following commutation relations~\cite{weinberg}:
%\begin{equation*}
% [J_i,J_j] = i\epsilon_{ijk}J_k \ \ [J_i,K_j] = i\epsilon_{ijk}K_k
%\end{equation*}
%\begin{equation*}
% [K_i,K_j] = -i\epsilon_{ijk}J_k \ \ [J_i,P_j] = i\epsilon_{ijk}P_k
%\end{equation*}
%\begin{equation*}
% [K_i,P_j] = -i\delta_{ij}H \  \ [K_i,H] = -iP_i
%\end{equation*}
%\begin{equation}\label{Palg2}
% [J_i,H] = [P_i, H] = [P_i, P_j] = 0.
%\end{equation}
The unitary representation of $\tilde{\mathcal{P}}^\uparrow_+$ is the direct sum of unitary irreducible representations (UIRs) on which the Casimir invariants of the Poincar\'e algebra act as multiples of the identity.  The two invariant operators are identified as the mass-squared $M^2 = P_\mu P^\mu$ and the negative square of the Pauli-Lubanski vector $W^2 = -w_\mu w^\mu$, where the four vector $w$ is
\begin{equation}
w_\mu = ({\bf P}\cdot{\bf J}, H{\bf J} - {\bf P}\times{\bf K})
\end{equation}
and $P^\mu = (H, {\bf P})$.
We will consider UIRs with positive definite mass labeled by $\ms =m^2$ and intrinsic spin $j$ such that for all vectors in the representation space $\phi\in\Phi(\ms,j)$,
\begin{equation}
M^2\phi = \ms\phi\ \mbox{and}\ W^2\phi = \ms j(j+1)\phi.
\end{equation}
The representation space $\Phi(\ms,j)$ is typically endowed with a topology given by the scalar product norm and is therefore the Hilbert space $\mathcal{H}(\ms,j)$.  Since we wish to use the non-normalizable basis vectors of momentum, technically we must work with a dense subspace $\Phi(\ms,j)\subset\mathcal{H}(\ms,j)$ with a stronger topology that allows for the nuclear spectral theorem and the continuity of the Poincar\'e algebra (for the general details of rigged Hilbert spaces or Gel'fand triplets, see \cite{bohmrhs}; for details relevant to the Poincar\'e algebra see \cite{rgv}). Since we are dealing with group representations (as opposed to, for example, semigroup representations) this will have no practical effect on the calculations other than giving the eigenstates of continuous observables a proper mathematical definition.

Within a particular UIR, a complete set of commuting operators is chosen of the form
\[
\{M^2, W^2, {\bf P}, \Sigma_3(P)\},
\]
where the operator $\Sigma_3(P)$ is a function of the generator of the Poincar\'e group and is constructed as
\begin{equation}\label{spinop}
\Sigma_\mu(\vP) = \sqrt{\ms}^{-1}U(\alpha(P))w_\mu U^{-1}(\alpha(P))= \sqrt{\ms}^{-1}\Lambda(\alpha(P))^\nu_{\ \mu}w_\nu.
\end{equation}
The operator $U(\alpha(P))$ (and its $4\times 4$ representation $\Lambda(\alpha(P))^\nu_{\ \mu}$ is an operator associated to a particular ``boost'' element $\alpha(p)$.  Since $[P_\mu, w_\nu]=0$, $P$ can be replaced with its eigenvalue $p$ when $\Sigma_3(P)$ acts on a momentum eigenvector.
The Poincar\'e group element $\alpha(p)$ has the property that it boosts the four momentum in the rest frame $p_R = (m, {\bf 0})$ to final momentum $p$, i.e.
\begin{equation}\label{boost}
\Lambda(\alpha(p))p_R = p.
\end{equation}
The choice of $\alpha(p)$ is not unique.  For $u\in\su$, the group
element $\alpha(p)u$ also fulfills (\ref{boost}).  The 4-momentum hyperboloid $p^2=\ms$ is isomorphic to the left coset space $Q = \slg/\su$ and the particular left coset $Q(p)=\{\alpha(p)\}$ contains all elements that satisfy (\ref{boost}).

Specifying which representative element $\alpha(p)$ of $Q(p)$ to use in (\ref{spinop}) gives different physical meanings for the spin component~\cite{kummer,kandpoly}.  Choosing $\alpha(p)$ to be $\ell(p)$, defined as
\begin{equation}\label{ell}
\ell(p) = {\left( \frac{\sigma^\mu p_\mu}{m} \right)}^{1/2} = \frac{\hat{m} + \sigma^\mu p_\mu}{[2m(m+ E_\ms(\mmp))]^{-1/2}},
\end{equation}
(where $\sigma^\mu = (1_2, {\bf \sigma})$, $m=\sqrt{\ms}$, $\hat{m} = m 1_2$, $p = (E_\ms(\mmp), \vp)$, $E_\ms(\mmp) = \sqrt{\ms + \mmp^2}$, and $\vp^2 = \mmp^2$) means that $\Lambda(\ell(p))=L(p)$ is the standard, rotation-free boost used in \cite{gingrich02,alsing02,ahn03,ahnarch03,soo04}.  Then we call $\Sigma_i(P)=S_i(P)$, and physically it is the $i$-th spin component in the particle rest frame.  

The choice  for CSCO $\{M^2, W^2, {\bf P}, S_3(P)\}$ leads to the Wigner 3-momentum spin basis for the expansion of the representation space $\Phi(\ms,j)$.  If $\phi\in\Phi(\ms,j)$ are chosen to be elements of the Schwartz space of ``well-behaved'' functions of the momentum, then improper eigenvectors $\kt{\wig}$, or Dirac eigenkets, of this CSCO  are elements of $\Phi^\times(\ms,j)$, the linear topological dual of $\Phi(\ms,j)$ and have the following properties:
\begin{eqnarray}
M^2\kt{\wig} &=& \ms \kt{\wig} \nonumber \\
W^2\kt{\wig} &=& \ms j(j+1)\kt{\wig}\nonumber \\
{\bf P}\kt{\wig} &=& {\bf p}\kt{\wig}\nonumber \\
S_3(P)\kt{\wig} &=& \chi\kt{\wig}.
\end{eqnarray}

Another choice for the representative element of the coset $Q(p)$ is $h(p)=\rho(p)\ell(p_z)$, where $p_z = (E_\ms(\mmp), 0, 0, \mmp)$ and $\rho(p)\in\su$ is the rotation that takes the 3-axis into the direction of $p$:
\begin{equation}
\rho(p) = e^{-i\phi\sigma^3/2}e^{-i\theta\sigma^2/2},
\end{equation}
where $\vp = \mmp(\sin\theta\cos\phi, \sin\theta\sin\phi, \cos\theta)$ and $\sigma^i$ are Pauli matrices.
Then spin operator is transformed into the helicity operator:
\begin{equation}
\Sigma_3(P) = H(P) = \frac{{\bf J}\cdot {\bf P}}{|{\bf P}|}.
\end{equation}
The CSCO $\{M^2, W^2, {\bf P}, H(P)\}$ leads to the helicity basis $\kt{\wigh}$, with the only difference is that
\begin{equation}
H(P)\kt{\wigh} = \lambda\kt{\wigh}.
\end{equation}
An advantage of the helicity basis is that $h(p)$ is well-defined even for massless particles whereas $\ell(p)$ is not.

One can convert between the helicity basis and the Wigner spin basis using
\begin{equation}
\kt{\wig} = \sum_\lambda D^j_{\lambda \chi}(\rho^{-1}(p))\kt{\wigh},
\end{equation}
where $D^j$ is the $2j+1$-th dimensional representation of the quantum mechanical rotation $\rho\in{\rm SU}(2)$.  Other bases are possible, such as the front-form basis (see \cite{kandpoly} for a review of this matter), but this work will focus on the Wigner spin and helicity bases.  Much of what is described in the rest of this section holds whether the Wigner spin basis or helicity basis is chosen and therefore we will use the notation $\xi$ to stand for either $\chi$ or $\lambda$.

Choosing a relativistically invariant normalization,
\begin{equation}\label{wignorm}
\bk{p',\xi'[\ms,j]}{\wigx} = 2 E(\mmp)\delta^3({\bf p'} - {\bf p})\delta_{\xi'\xi},
\end{equation}
gives the following form to the expansion of a vector $\phi\in\Phi(\ms,j)$ with invariant measure
\begin{equation}
\phi = \sum_\xi \int \frac{d^3{\bf p}}{2 E(\mmp)} \kt{\wigx}\bk{\wigx}{\phi}.
\end{equation}
For $\phi\in\Phi(\ms,j)$ and $\kt{\wig}\in\Phi^\times(\ms,j)$ (the topological dual of $\Phi(\ms,j)$), the Poincar\'e transformations then are represented in this infinite dimensional basis as
\begin{subequations}\label{wigtrans}
\begin{eqnarray}
U(\alpha, a)\phi_\chi( p) &=& \br{\wig}U(\alpha, a)\kt{\phi}\nonumber \\
&=& e^{ip\cdot a}\sum_{\chi'}D_{\chi'\chi}^j(W(\alpha, \Lambda^{-1}(\alpha)p))\phi_{\chi'}( \Lambda^{-1}(\alpha)p)
\end{eqnarray}
or equivalently 
\begin{equation}
U(\alpha, a)\kt{\wig} = e^{-i\Lambda(\alpha)p\cdot a}\sum_{\chi'}D_{\chi'\chi}^{j}(W(\alpha, p)) \kt{\Lambda(\alpha)p,\chi'[\ms,j]},
\end{equation}
\end{subequations}
where $W(\alpha, p)\in{\rm SU}(2)$ is the Wigner rotation.  The Wigner rotation depends on the choice of representant $\alpha(p)$:  
\begin{equation}\label{wignerrot}
W(\alpha, p) = \alpha^{-1}(\Lambda(\alpha)p)\alpha\alpha(p).
\end{equation}
where $\alpha(p) = \ell(p)$ in the Wigner spin basis and $\alpha(p)=h(p)$ in the helicty basis.  For clarity, we will sometimes denote these two different Wigner rotations associated with the same Lorentz transformation and momentum but different boosts as $W_\ell(\alpha,p)$ and $W_h(\alpha,p)$.  In \cite{gingrich02}, it was correctly noted that for $u\in\su$
\begin{equation}
W_\ell(u,p) = 1,
\end{equation}
which is one advantage of the Wigner spin basis.  However, it is not correct to say that the transformation (\ref{wigtrans}) associated with a pure rotation does not depend on the momentum, because the momentum is implicitly included in the definition of the spin component. 

Finally, we note that in non-relativistic quantum mechanics, one can write basis vectors of momentum and spin as direct products like 
\begin{equation}\label{nonrel}
\kt{\vp}\otimes\kt{\xi}
\end{equation}
and these transform under the group of Galilean transformations.  Galilean boosts and space-time translations act only on the momentum degrees of freedom whereas rotations act on both the momentum space and spin space.  Rotations are not complicated by the momentum dependence of the spin definition and are implemented as
\begin{equation}
U(\rho)\kt{\vp}\otimes\kt{\xi} = \kt{R(\rho)\vp}\otimes\sum_{\xi'}D^j_{\xi'\xi}(\rho)\kt{\xi},
\end{equation}
where $\rho\in\su$ and $R(\rho)\in\mathrm{SO}(3)$.  When considering UIRs of $\ptr$, not only is $\rho$ in the rotation representation matrix changed to a Wigner rotation, but the separation (\ref{nonrel}) makes no sense because of the implicit momentum-dependence of the spin component.  See \cite{ahn03,ahnarch03,kim04} for examples where this (erroneous) direct product assumption is applied.

\section{Relativistic Two-Particle Bases and Clebsch-Gordan Coefficients for the Poincar\'e Group}

The state space of a multi-particle system can be decomposed into a direct sum of irreducible representations of $\ptr$.  For example, for the direct product two UIRs associated to particles with masses $m_i$ and spins $j_i$, we have~\cite{schaaf}
\begin{equation}\label{dsdecom}
\Phi(m_1^2, j_1)\otimes\Phi(m_2^2, j_2) = \sum_{j = j_0}^\infty \int_{(m_1 + m_2)^2}^\infty d\mu(\ms) \sum_\eta \Phi(\ms,j)_\eta.
\end{equation}
The sum over total angular momentum $j$ begins at $j_0=0$ if both particles are fermions or bosons and at $j_0=1/2$ for an unlike pair and the integral is a direct integral over center-of-mass energy squared $\ms$ with measure $d\mu(\ms)$.   As we will discuss in more detail below, a particular UIR $\Phi(\ms,j)$ may appear in the direct sum decomposition multiple times and $\eta$ labels this degeneracy.

Poincar\'e covariance of the interaction assures that any dynamical entanglement will be constrained to UIRs in the direct sum with particular values for $\ms$ and $j$.  All transitions amplitudes are diagonal in $\ms$ and $j$ because even the interacting generators should satisfy the same commutation rules as the non-interacting Poincar\'e algebra.  Therefore, in this section we will use the Clebsch-Gordan coefficients for $\ptr$ to decompose two particle spaces into UIRs of the total generators and thereby obtain set of entangled basis states with fixed $\ms$ and $j$ that do not mix under any Poincar\'e transformation.

The CGCs transform from the direct product basis to the direct sum basis.
We make the following choice of notation and normalization for the two-particle direct product states valid for either the Wigner spin basis or helicty basis (the single-particle invariants have been suppressed in the equations below):
\begin{eqnarray}\label{normdp}
\bk{1 \otimes 2}{1' \otimes 2'} &=& \bk{p_1 \xi_1 ;p_2 \xi_2 }{p'_1 \xi'_1 ;p'_2 \xi'_2 }\nonumber\\
&=& 2E_1(\mmp_1) 2E_2(\mmp_2)  \delta^3(\vp_1 - \vp_1')\delta^3(\vp_2 - \vp_2')\delta_{\xi_1 \xi'_1}\delta_{\xi_2 \xi'_2}.
\end{eqnarray}
On the direct product vectors, the representation of $\ptr$ on $\Phi(\ms_1,j_1) \otimes \Phi(\ms_2,j_2)$ (and its extension to $\Phi^\times(\ms_1,j_1) \otimes \Phi^\times(\ms_2,j_2)$) is the direct product of the one-particle transformation representations (\ref{wigtrans}):
\begin{equation}
U(\alpha,a)\kt{1\otimes 2} = U_1(\alpha,a)\kt{p_1 \xi_1 }\otimes  U_2(\alpha,a)\kt{p_2 \xi_2 }.
\end{equation}
The CSCO for this basis is the sum of the one-particles CSCOs
\[
\{M_1^2, W_1^2, {\bf P}_1, \Sigma_3(P_1)_1, M_2^2, W_2^2, {\bf P}_2, \Sigma_3(P_2)_2\},
\]
where, for example, the notation $M_1^2$ implies the natural extension to the direct product space, $M_1^2 \otimes I$ and where $\Sigma_3$ could mean either the Wigner spin-operator or the helicity operator.

There are many possible choices of CSCO for the direct sum basis, but all have a form like
\begin{equation}
\{M^2, W^2, {\bf P}, \Sigma_3(P), \eta^{(\mathrm{op})}_1, \eta^{(\mathrm{op})}_2, M_1^2, W_1^2, M_2^2, W_2^2\}
\end{equation}
The Poincar\'e algebra from the total system is constructed from the sums of the generators of the single particle generators, i.e. $P^\mu = P_1^\mu \otimes I + I \otimes P^\mu_2$.  The total operators $M^2$ and $W^2$ are again invariants specifying how transformation act in the UIR, and the single particle invariants $M_i^2$ and $W_i^2$ are also invariants of the total system, and so the direct sum space $\Phi(\ms,j)_\eta$ implicitly also carries the labels of the one-particle invariants.  Additionally, there are two more operators, $\eta^{(\mathrm{op})}_i$, that label the degeneracy of the UIR in the direct sum.  They depend on the choice made for the degeneracy parameters $\eta_i$ (see below).  Finally, there are the non-invariant momentum and spin-component operators labeling the states within the UIR space $\Phi(\ms,j)_\eta$.

We choose the normalization eigenkets of the direct sum CSCO in the following fashion,
\begin{eqnarray}\label{normds}
\bk{\tau}{\tau'} &=& \bk{p \xi [\ms j \eta_1 \eta_2]}{p' \chi' [\ms' j' \eta'_1 \eta'_2]} \nonumber\\
&=& 2 E_\ms(\mmp)\ms \delta^3(\vp - \vp')\delta_{\xi \xi'}\delta(\ms - \ms')\delta_{jj'}\delta_{\eta_1\eta'_1}\delta_{\eta_2\eta'_2}.
\end{eqnarray}
These kets obey the same transformation rule (\ref{wigtrans}) as the one-particle UIRs:
\begin{equation}
U(\alpha, a)\kt{p \xi [\ms j \eta_1 \eta_2]} = e^{-i\Lambda(\alpha)p\cdot a}\sum_{\xi'}D_{\xi'\xi}^{j}(W(\alpha, p)) \kt{\Lambda(\alpha)p \xi' [\ms j \eta_1 \eta_2]}.
\end{equation}

The CGCs for the quantum mechanical Poincar\'e group then are the amplitudes
\[
\bk{1\otimes 2}{\tau} =\bk{p_1 \xi_1 p_2 \xi_2}{p \chi [\ms j \eta_1 \eta_2]}.
\]
Their structure clearly depends on how the one-particle UIRs are constructed (and therewith the one-particle CSCOs) as well as the choice of coupling scheme.  A general scheme for constructing the CGCs of $\ptr$ is the double-coset method~\cite{mackey} used in \cite{moussa,schaaf,klinksmith}.  Whippman~\cite{whippman} uses a nice alternative technique involving group integration over representation matrix elements.

The CGCs for the direct product of two, distinguishable representations of $\ptr$ can be split into a kinematics/normalization term and an angular correlation term:
\begin{equation}\label{cgcres}
\bk{1\otimes 2}{\tau} = K_{12}(p_1 p_2 ; p)A_{12}(p_1 \xi_1 p_2 \xi_2; p \xi j \eta_1 \eta_2).
\end{equation}
The term $K_{12}$ is the kinematic term involving momentum conservation and will have the same form in the any coupling scheme.  It depends on the normalizations (\ref{normdp}) and (\ref{normds}) and looks like:
\begin{equation}\label{kin}
K_{12}(p_1 p_2 ; p) = \frac{2\sqrt{2}}{\Delta(\ms, \ms_1, \ms_2)^{1/4}}    \ms\, 2E_\ms(\mmp)  \delta^3(\vp_1 + \vp_2 - \vp)\delta((p_1 + p_2)^2 - \ms),
\end{equation}
where
\[
\Delta(\ms, \ms_1, \ms_2) = \ms^2 + \ms_1^2 + \ms_2^2 - 2(\ms\ms_1 + \ms\ms_2 + \ms_1\ms_2).
\]
Note that the magnitude of the center-of-mass momentum of both particles is 
\[
\mk = \sqrt{\frac{\Delta(\ms, \ms_1, \ms_2)}{4\ms}}.
\]
A choice of phase convention has been made such that $K=K^*$.
The term $A_{12}$ contains the information about the angular distribution and spin correlations and differs depending on the choice of boost/spin component. 

Working with the Wigner spin basis, the degeneracy can be labeled according to the spin-orbit coupling scheme of Joos~\cite{joos} and MacFarlane~\cite{macfarlane}.  In this scheme, the single particle generators can be combined to form total intrinsic spin ${\bf S}$ and orbital angular momentum ${\bf L}$ operators, and $\eta = \{s,l\}$ are their eigenvalues.  This choice has the advantage that it is familiar from non-relativistic quantum mechanics and that, if the full Poincar\'e group $\pt$ including discrete transformations are allowed, there is a simple correspondence between $\eta$ and the eigenvalues of parity and charge parity for the composite system.

In the helicity basis, important conventions were established by Jacob and Wick~\cite{jacobwick} and have a long history of use in partial wave analysis of scattering experiments.  In the helicity coupling, the $\eta$ are the eigenvalues of the two single particle helicity operators in the rest frame of the system.  We will consider the CGCs associated with each choice of direct sum basis.

\subsection{Wigner Basis Spin-Orbit Coupling CGCs}

In this basis, the CSCO is
\begin{equation}
\{M^2, W^2, {\bf P}, S_3(P), {\bf L}^2, {\bf S}^2, M_1^2, W_1^2, M_2^2, W_2^2\},
\end{equation}
where $L$ is the orbital angular momentum and $S$ is the coupled intrinsic spin (see \cite{kandpoly}, p.~329 for an example of their explicit constructions in terms of the one-particle generators).

 For spin orbit-coupling, the angular term is~\cite{macfarlane,whippman}
\begin{eqnarray}\label{ang}
A_{12}(p_1 \chi_1 p_2 \chi_2; p \chi j l s) &=& \sum_{\chi'_1 \chi'_2} D^{j_1}_{\chi'_1 \chi_1}(u(p, p_1)) D^{j_2}_{\chi'_2 \chi_2}(u(p, p_2))\nonumber\\
&&\times
\sum_{l_3 s_3} C(s j_1 j_2; s_3 \chi'_1 \chi'_2) C(j l s; \chi l_3 s_3)
(-)^\chi Y_{ll_3}(\hat{\bf \Omega}(p_1,p_2)),
\end{eqnarray}
where the following conventions and notations have been chosen:
\begin{itemize}
\item The rotation $W_\ell(\ell^{-1}(p),p_i)$ when applied to a single particle ket $\kt{p_i, \chi_i}$ effects the mixing of the spin coordinates when the basis vector for particle $i$ is transformed into COM frame in which the spin coupling takes place.  The argument $u(p, p_i)$ of the rotation matrix $D^{j_i}$ is then the inverse of this Wigner rotation:
\begin{equation}\label{invwig}
u(p,p_i) = W_\ell^{-1}(\ell^{-1}(p),p_i).
\end{equation}

\item The CGC's for the rotation group are chosen according to standard phase conventions (i.e., they are all real).  They are
\begin{equation}
C(j j_1 j_2; \chi \chi_1 \chi_2) = \bk{\chi_1 \chi_2 [j_1 j_2]}{\chi [j j_1 j_2]}
\end{equation}
where $\chi = \chi_1 + \chi_2$.  These are used to couple the spins of the two particles and to couple the total spin with the orbital angular momentum.
\item The spherical harmonic $Y_{ll_3}(\hat{\bf \Omega})$ describes the angular dependence on the orbital angular momentum and is a function of the unit-normalized relative momentum $e= (0, \hat{\bf \Omega})$ in the barycentric frame:
\begin{equation}\label{sphe}
e(p_1,p_2) = \left(\frac{\ms}{\Delta(\ms,\ms_1,\ms_2)}\right)^{1/2} L^{-1}(p_1 + p_2)\{p_1 - p_2 - [(\ms_1 - \ms_2)/\ms](p_1 + p_2)\}.
\end{equation}
\item The phase factor $(-)^\chi$ is introduced for so that the direct sum basis vectors transform in the usual way under time reversal~\cite{cgc}.
\end{itemize}

Each vector in $\Phi(\ms, j)_{ls}$ describes a particular kind of two-particle entanglement and the basis vectors of $\Phi(\ms, j)_{ls}$ span a space of entangled two-particle states that is invariant under Poincar\'e transformations.  This entanglement involves both spin and momentum.  An additional property of spin-orbit coupling scheme is that the spaces $\Phi(\ms, j)_{ls}$ are eigenspaces of the parity and charge parity operator~\cite{cgc}.  

\subsection{Helicty Basis Coupling CGCs}

In this basis, the CSCO is
\begin{equation}
\{M^2, W^2, {\bf P}, H(P), H^{cm}_1, H^{cm}_2, M_1^2, W_1^2, M_2^2, W_2^2\},
\end{equation}
where $H^{cm}_i$ is the helicity of the $i$-th particle in the center-of-mass frame and is invariant.  Its eigenvalues are $\tilde{\lambda}_i$.

For helicity-coupling~\cite{jacobwick,schaaf,whippman}, the angular term has the form
\begin{eqnarray}\label{angh}
A_{12}(p_1 \la_1 p_2 \la_2; p \la j \tl_1 \tl_2) &=& \left(\frac{2j + 1}{4\pi}\right)^{1/2} \sum_{\tl_1 \tl_2} D^{j_1}_{\tl_1 \la_1}(u(p, p_1)) D^{j_2}_{\tl_2 \la_2}(u(p, p_2))\nonumber\\
&&\times
D^{j}_{\la (\tl_1 - \tl_2)}(\rho(\hat{\bf \Omega}_1)),
\end{eqnarray}
where $u(p,p_1)$ is the inverse Wigner rotation (\ref{invwig}), but this time calculated using the helicity boost $h(p)$.  The rotation $\rho(\hat{\bf \Omega}_1)$ is the rotation that performs a rotation of $-\phi_1$ around the $z$-axis and then aligns the $z$-axis with the direction $\hat{\bf \Omega}_1$; in other words, using the standard $zyz$ Euler angle form and the Pauli matrices $\sigma^i$~\cite{rose}:
\begin{equation}
\rho(\hat{\bf \Omega}_1)= e^{-i\phi_1 \sigma^3/2}e^{-i\theta_1 \sigma^2/2}e^{i\phi_1 \sigma^3/2},
\end{equation}
and so
\begin{equation}
D^{j}_{\la (\tl_1 - \tl_2)}(\rho(\hat{\bf \Omega}_1)) = e^{-i\la\phi_1}d^j_{\la (\tl_1 - \tl_2)}(\theta_1)e^{-i(\tl_1 - \tl_2)\phi_1}.
\end{equation}

The helicity coupling scheme does not produce eigenspaces of the discrete symmetries, but it is generalizable to massless particles and more easily generalized to multiparticle ($N>2$) direct products~\cite{werhle,klinksmith,kummer,klink92}.

\section{Basis vectors for dynamically entangled decay products}

For the sake of specificity, consider the entanglement of two particles due to the decay of a parent particle of mass $M$ and spin $j$.
Far from the decay site, there will be no interaction between the decay products and the state space $\Phi_{12}$ will be a subspace of the direct product of the UIRs associated with each daughter particle $\Phi(m_1^2,j_1)\otimes\Phi(m_2^2,j_2)\supset\Phi_{12}$.  However, because of the Poincar\'e covariance of the dynamics, all elements of $\Phi_{12}$ must have a COM energy squared $\ms=M^2$ and total angular momentum $j$, i.e.,
\begin{equation}
\Phi_{12} = \bigoplus_{\eta\in d(j)}\Phi(M^2, j)_\eta,
\end{equation}
Further information about the dynamics may restrict the direct sum to a single value of $\eta$.  For example, in the spin-orbit coupling scheme, $\eta$ determines the overall parity and charge parity of the space (see below).

The full mathematical details and explanation of the phenomenological signatures of relativistic decay processes can be described using the relativistic Gamow vector~\cite{rgv}, an element of an irreducible representation of the Poincar\'e semigroup.  Since we are just interested in the kinematic correlations of the decay products, we can gloss over most of the details of mathematical rigor without changing any conclusions.

In what follows, we will explore these basis vectors for a simple case of a particle/antiparticle pair with mass $m_i=m$ and spin $j_i=1/2$.  In particular, we will look at the case that the parent particle  has spin either $j=0$ or $j=1$ and find the basis vectors for the direct sum UIRs.

First we consider the spin-orbit coupling scheme for Wigner spin vectors.
Simplifying to the COM reference frame, the rest vector of the composite states $\kt{p_R \chi [\ms j l s]}\in\Phi^\times(\ms j)_{ls}$ can be expressed using the CGCs in the following form:
\begin{eqnarray}\label{com}
\kt{p_R \chi [\ms j l s]} &=& \sum_{\chi_1,\chi_2}\int\frac{d^3\vp_1d^3\vp_2}{4 E_1(\mmp_1)E_2(\mmp_2)} \kt{p_1 \chi_1 ;p_2 \chi_2 }K_{12}(p_1 p_2 ; p_R)A_{12}(p_1 \chi_1 p_2 \chi_2; p_R \chi j \eta)\nonumber\\
&=& \frac{\sqrt{2}}{2}\Delta^{1/4}(\ms, \ms_1, \ms_2)\sum_{\chi_1,\chi_2} \int d^2\hat{\bf \Omega} A_{12}(\tilde{p}_1 \chi_1 \tilde{p}_2 \chi_2; p_R \chi j l s)
\kt{\tilde{p}_1 \chi_1;\tilde{p}_2 \chi_2},
\end{eqnarray}
where $\tilde{p}_1 = (E_1(\mk), \mk\hat{\bf \Omega})$ and $\tilde{p}_2 = (E_2(\mk), -\mk\hat{\bf \Omega})$.
The factor $A_{12}$ takes the simpler form
\begin{equation}
A_{12}(\tilde{p}_1 \chi_1 \tilde{p}_2 \chi_2; p_R \chi j l s) = 
\sum_{l_3 s_3} C(s j_1 j_2; s_3 \chi_1 \chi_2) C(j l s; \chi l_3 s_3)
(-)^\chi Y_{ll_3}(\hat{\bf \Omega})
\end{equation}
in the COM frame since $\ell(p_R) = I$.  Note that even in the COM frame there is an implicit momentum dependence in $A_{12}$ because of the spin components.

The possible values for $l$ and $s$ are implicit in the rotation group CGCs, $s\in\{0,1\}$ and $|l-s|<j<l+s$.  In this case, the spaces $\Phi(\ms, j)_{ls}$ are eigenspaces of the parity and charge parity operator~\cite{cgc} with the parity $\pi_P = \pi_{P1}\pi_{P2} (-1)^l = (-1)^{l +1}$ (since the parity of fermions and antifermions are opposite) and charge parity $\xi_C = (-1)^{l +s}$.  These six relevant UIRs $\Phi(\ms, j)_{ls}$ are summarized in Table I.

\begin{table}

\begin{tabular}{|c|c|c|c|c|}
\hline
$j$ & $s$ & $l$ & $\pi_P$ & $\xi_C$ \\ \hline
$0$ & $0$ & $0$ & $-$ & $+$\\ \cline{2-5}
    & $1$ & $1$ & $+$ & $+$\\ \hline
$1$ & $0$ & $1$ & $+$ & $-$\\ \cline{2-5}
    & $1$ & $0$ & $-$ & $-$\\ \cline{2-5}
    & $1$ & $1$ & $+$ & $-$\\ \cline{2-5}
    & $1$ & $2$ & $-$ & $+$\\ \hline
\end{tabular}

\caption{Possible values of $l$ and $s$ and assignment of parity and charge parity eigenvalues with $j=0$ and $j=1$ for UIRs appearing in the direct sum decomposition of a spin 1/2 fermion/antifermion pair.}
\end{table} 

For the two $j=0$ cases, the COM frame subspace of $\Phi(\ms, j)_{ls}$ is spanned by a single vector:
\begin{eqnarray}
\kt{p_R \chi [\ms 0 0 0]} &=& \sqrt{\frac{1}{8\pi}}\Delta^{1/4}\int d^2\hat{\bf \Omega} \left(
\kt{\tilde{p}_1 1/2;\tilde{p}_2 -1/2} - \kt{\tilde{p}_1 -1/2;\tilde{p}_2 1/2}\right)\ \mbox{or}\\
\kt{p_R \chi [\ms 0 1 1]} &=& \sqrt{\frac{1}{32\pi}}\Delta^{1/4}\int d^2\hat{\bf \Omega} \left(
\sin\theta e^{-i\phi}\kt{\tilde{p}_1 1/2;\tilde{p}_2 1/2} - \cos\theta\kt{\tilde{p}_1 1/2;\tilde{p}_2 -1/2}\right.\nonumber\\
&& \left. -   \cos\theta\kt{\tilde{p}_1 -1/2;\tilde{p}_2 +1/2} - \sin\theta e^{i\phi}\kt{\tilde{p}_1 -1/2;\tilde{p}_2 -1/2}\right),
\end{eqnarray}
where $\hat{\bf {\Omega}} = (\sin\theta\cos\phi, \sin\theta\sin\phi, \cos\theta)$.  The $(ls)=(00)$ case looks like the typical spin singlet case (except for the momentum direction integral), but the $(11)$ is quite different and shows explicitly the momentum-dependence of the spin correlations.  Since the definition of the spin component involves momentum, separating out momentum- and spin-correlations may not be meaningful.  Future work should decide whether such a division can be measured and is meaningful, and results on the ill-defined nature of spin entropy in relativistic systems~\cite{peres02,czachor03} suggest it will not be analogous to the non-relativistic case.

In Table II, we have included the angular coefficients of the twelve basis vectors for the four $j=1$ UIRs.  Rotations will create superpositions of these basis states, but not Poincar\'e transformation will mix different $l$ and $s$ values.  These vectors form sensible bases for dynamically-entangled states.

\begin{table}

\begin{tabular}{|c|c|r|c|c|c|c|}
\hline
$s$ & $l$ & $\chi$& $\kt{\tilde{p}_1 1/2;\tilde{p}_2 1/2}$ & $\kt{\tilde{p}_1 1/2;\tilde{p}_2 -1/2}$ & $\kt{\tilde{p}_1 -1/2;\tilde{p}_2 1/2}$ & $\kt{\tilde{p}_1 -1/2;\tilde{p}_2 -1/2}$\\ \hline
$0$ & $1$ & $1$ & $0$ & $\sqrt{\frac{3}{16\pi}}\sin\theta e^{i\phi}$ & $-\sqrt{\frac{3}{16\pi}}\sin\theta e^{i\phi}$ & $0$\\ \cline{3-7}
        & & $0$ & $0$ & $\sqrt{\frac{3}{8\pi}}\cos\theta $& $-\sqrt{\frac{3}{8\pi}}\cos\theta$ & $0$\\ \cline{3-7}
        & & $-1$ & $0$ & $-\sqrt{\frac{3}{16\pi}}\sin\theta e^{-i\phi} $& $\sqrt{\frac{3}{16\pi}}\sin\theta e^{-i\phi}$ & $0$\\ \hline
$1$ & $0$ & $1$ & $\sqrt{\frac{1}{4\pi}}$ & $0$ & $0$ & $0$\\ \cline{3-7}
        & & $0$ & $0$ & $\sqrt{\frac{1}{8\pi}}$& $\sqrt{\frac{1}{8\pi}}$ & $0$\\ \cline{3-7}
        & & $-1$ & $0$ & $0$& $0$ & $-\sqrt{\frac{1}{4\pi}}$\\ \hline
$1$ & $1$ & $1$ & $\sqrt{\frac{3}{8\pi}}\cos\theta$ & $\sqrt{\frac{3}{16\pi}}\sin\theta e^{i\phi}$& $\sqrt{\frac{3}{16\pi}}\sin\theta e^{i\phi}$ & $0$\\ \cline{3-7}
        & & $0$ & $-\sqrt{\frac{3}{16\pi}}\sin\theta e^{-i\phi}$ & $0$ & $0$ & $-\sqrt{\frac{3}{16\pi}}\sin\theta e^{i\phi}$\\ \cline{3-7}
        & & $-1$ & $0$ & $\sqrt{\frac{3}{16\pi}}\sin\theta e^{-i\phi}$& $\sqrt{\frac{3}{16\pi}}\sin\theta e^{-i\phi}$ & $-\sqrt{\frac{3}{8\pi}}\cos\theta$\\ \hline
$1$ & $2$ & $1$ & $-\sqrt{\frac{1}{8\pi}}(\frac{3}{2}\cos^2\theta - \frac{1}{2})$ & $-\sqrt{\frac{9}{32\pi}}\sin\theta\cos\theta e^{i\phi}$ & $-\sqrt{\frac{9}{32\pi}}\sin\theta\cos\theta e^{i\phi}$ & $-\sqrt{\frac{9}{32\pi}}\sin^2\theta e^{2i\phi}$\\ \cline{3-7}
        & & $0$ & $\sqrt{\frac{9}{16\pi}}\sin\theta\cos\theta e^{-i\phi}$ & $-\sqrt{\frac{1}{4\pi}}(\frac{3}{2}\cos^2\theta - \frac{1}{2})$ & $-\sqrt{\frac{1}{4\pi}}(\frac{3}{2}\cos^2\theta - \frac{1}{2})$ & $-\sqrt{\frac{9}{16\pi}}\sin\theta\cos\theta e^{i\phi}$\\ \cline{3-7}
        & & $-1$ & $-\sqrt{\frac{9}{32\pi}}\sin^2\theta e^{-2i\phi}$ & $\sqrt{\frac{9}{32\pi}}\sin\theta\cos\theta e^{-i\phi}$ & $\sqrt{\frac{9}{32\pi}}\sin\theta\cos\theta e^{-i\phi}$ & $-\sqrt{\frac{1}{8\pi}}(\frac{3}{2}\cos^2\theta - \frac{1}{2})$\\ \hline
\end{tabular}
\caption{Angular part of CGCs for Wigner spin basis using spin-orbit coupling for $j=1$.}
\end{table}

Entangled bases can also be constructed with the helicity basis and helicity coupling.
In the center-of-mass frame (\ref{angh}) simplifies considerably since this coupling scheme relies on the helicities of the component particles in the COM frame:
\begin{equation}
A_{12}(\tilde{p}_1 \tl_1 \tilde{p}_2 \tl_2; p_R \la j \tl_1 \tl_2) = \left(\frac{2j + 1}{4\pi}\right)^{1/2} 
D^{j}_{\la (\tl_1 - \tl_2)}(\rho(\hat{\bf \Omega}_1)).
\end{equation}
Using the analogous result from  spin-orbit coupling in the COM frame (\ref{com}), we have
\begin{eqnarray}\label{hvec}
\kt{p_R \la [\ms j \tl_1 \tl_2]} &=& \int\frac{d^3\vp_1d^3\vp_2}{4 E_1(\mmp_1)E_2(\mmp_2)} \kt{p_1 \tl_1 ;p_2 \tl_2}K_{12}(p_1 p_2 ; p_R)A_{12}(p_1 \tl_1 p_2 \tl_2; p_R \la j \tl_1 \tl_2)\nonumber\\
&=& \frac{\sqrt{2}}{2}\Delta^{1/4}\int d^2\hat{\bf \Omega} A_{12}(\tilde{p}_1 \tl_1 \tilde{p}_2 \tl_2; p_R \la j \tl_1 \tl_2)
\kt{\tilde{p}_1 \tl_1;\tilde{p}_2 \tl_2}.
\end{eqnarray}
The angular part of the CGCs are given in Table III.  The vectors of (\ref{hvec}) using these angular coefficients are not entangled, except in the most trival sense of opposite momenta.  However, any superposition of the vectors of (\ref{hvec}) for a given $j$ value will also show spin correlations.  A typical decay process will lead to such a superposition, and therefore entanglement.

\begin{table}

\begin{tabular}{|c|c|c|r|c|}
\hline
$\tl_1$ & $\tl_2$ & $j$ & $\la$ & $A_{12}$ \\
\hline
$+1/2$ & $+1/2$ & $0$ & $0$ & $\sqrt{1/4\pi}$ \\ \hline
$-1/2$ & $-1/2$ & $0$ & $0$ & $\sqrt{1/4\pi}$ \\ \hline
$+1/2$ & $+1/2$ & $1$ & $1$ & $-\sqrt{3/8\pi}e^{-i\phi}\sin\theta$ \\ \cline{4-5}
& & & 									$0$ & $\sqrt{3/4\pi\cos\theta}$ \\ \cline{4-5}
& & & 									$-1$ & $\sqrt{3/8\pi}e^{i\phi}\sin\theta$ \\ \hline
$+1/2$ & $-1/2$ & $1$ & $1$ & $\sqrt{3/16\pi}(1 + \cos\theta)$ \\ \cline{4-5}
& & & 									$0$ & $\sqrt{3/8\pi}e^{i\phi}\sin\theta$ \\ \cline{4-5}
& & & 									$-1$ & $\sqrt{3/16\pi}e^{2i\phi}(1 - \cos\theta)$ \\ \hline
$-1/2$ & $+1/2$ & $1$ & $1$ & $\sqrt{3/16\pi}e^{-2i\phi}(1 - \cos\theta)$ \\ \cline{4-5}
& & & 									$0$ & $-\sqrt{3/8\pi}e^{-i\phi}\sin\theta$ \\ \cline{4-5}
& & & 									$-1$ & $\sqrt{3/16\pi}(1 + \cos\theta)$ \\ \hline
$-1/2$ & $-1/2$ & $1$ & $1$ & $-\sqrt{3/8\pi}e^{-i\phi}\sin\theta$ \\ \cline{4-5}
& & & 									$0$ & $\sqrt{3/4\pi\cos\theta}$ \\ \cline{4-5}
& & & 									$-1$ & $\sqrt{3/8\pi}e^{i\phi}\sin\theta$ \\ \hline
\end{tabular}
\caption{Angular part of CGCs for helicity basis using helicity coupling for $j=0$ and $j=1$.}
\end{table}

\section{Conclusion}

Constructions of entangled basis vectors are the first step in calculations of Bell-type inequalities and their properties under Lorentz transformations.  Working with vectors like  (\ref{crap}), a conclusion of \cite{alsing02} is that entanglement fidelity is preserved, whereas \cite{ahn03,ahnarch03} claim that entanglement is not invariant under Lorentz boosts.  The authors of \cite{czachor97} and \cite{ahn03,ahnarch03} use different methods to come to the conclusion that Lorentz boosts reduce the amount of spin correlation and that entangled bi-partite systems may appear to satisfy Bell's inequality to a highly relativistic observer, whereas \cite{terashima03} show that perfect (anti-)correlations still appear if the correct measurements of spin are made.

Some of this difference in opinion stems from a mistake made by the authors of \cite{ahn03,ahnarch03} in expressing the state of a particle as the direct product of the momentum state and the spin state, as discussed above.  However, another obscuring issue, which this paper hopes to make clear, is that states like (\ref{crap}) do not have the kinematic correlations that arise from dynamic entanglement.  For analysis of actual scattering and decay experiments, basis vectors constructed using the CGCs of $\ptr$ have proved their usefulness many times, and it should be no surprise that similar techniques will be required for relativistic quantum information theory.

There are some results about entanglement measures and spin entropy of these kinds of relativistic entangled states (see \cite{peres04} for a review), but there are many more unanswered questions.  Future research by this author will present a rigorous solution to the properties of Bell's inequalities under Poincar\'e transformations and will consider appropriate measures for entanglement in systems like these.

\begin{acknowledgments}
The author would like to thank Y.S.~Kim, organizer of the ``Second Feynman Festival,'' held at University of Maryland, College Park from 20-26 August, 2004.  I was made aware of this problem at that conference thanks to a talk given by Doyeol Ahn.
\end{acknowledgments}

\end{document}